# Anisotropic buckling of few-layer black phosphorus

Luis Vaquero-Garzon[a], Riccardo Frisenda*[a], Andres Castellanos-Gomez*[a]

[a.] *Materials Science Factory. Instituto de Ciencia de Materiales de Madrid (ICMM-CSIC), Madrid, E-28049, Spain.*

* Corresponding authors: riccardo.frisenda@csic.es and andres.castellanos@csic.es

**When a two-dimensional material, adhered onto a compliant substrate, is subjected to compression it can undertake a buckling instability yielding to a periodic rippling. Interestingly, when black phosphorus flakes are compressed along the zig-zag crystal direction the flake buckles forming ripples with a 40% longer period than that obtained when the compression is applied along the armchair direction. This anisotropic buckling stems from the puckered honeycomb crystal structure of black phosphorus and a quantitative analysis of the ripple period allows us to determine the Youngs's modulus of few-layer black phosphorus along the armchair direction ($E_{bP\_AC}$ = 35.1 ± 6.3 GPa) and the zig-zag direction ($E_{bP\_ZZ}$ = 93.3 ± 21.8 GPa).**

Since its isolation in 2014,[1–6] few-layer black phosphorus (bP) keeps attracting the interest of scientific community because of its remarkable electronic (*i.e.* ultrahigh charge carrier mobility, ambipolar field effect, etc)[7–10] and optical (*i.e.* narrow direct gap, strong quantum confinement effect, large band gap electrical tunability, etc)[11–20] that has motivated its application in many electronic and optoelectronic devices.[5,21–23] Strikingly, although the electronic and optical properties have been thoroughly characterized its mechanical properties, that have a crucial role in its applicability in flexible electronics and nanoelectromechanical systems, have been barely studied experimentally and these works do not provide a good consensus in their results, especially in the value of the elastic modulus of black phosphorus along the zig-zag direction.[24–28] One possible explanation for the scattering of results in the literature is the environmental instability of black phosphorus: few-layer black phosphorus flakes degrade upon atmospheric exposure within hours.[29–38] Favron *et al*. and Zhou *et al*. have deduced that the degradation most likely occurs as a result of photo-induced oxidation, forming phosphorus oxide species, from oxygen absorbed in the accumulated water at the surface of exfoliated flakes exposed to ambient conditions.[32,39,40] In the previous works, the mechanical testing methods used require exposing the flakes to air for relatively long periods of time and in some cases the studied flakes have been even subjected to several wet-chemistry microfabrication steps[28] or exposed to electron beam irradiation.[24] Therefore, there is a need for a technique that allows to study the intrinsic mechanical properties of pristine black phosphorus flakes right after their exfoliation.

Here, we have studied the mechanical properties of few-layer black phosphorus flakes by buckling induced metrology,[41] which has been recently demonstrated to be a very fast and reliable way to measure the Young's modulus of thin films,[41–43] organic semiconductors[44] and 2D materials.[45–49] Interestingly, when the black phosphorus flakes are subjected to uniaxial compression, they tend to buckle forming ripples predominantly parallel to the zig-zag (referred to as ZZ hereafter) crystal axis of the black phosphorus lattice, allowing us to determine the Young's modulus along the arm-chair (referred to as AC hereafter) direction. Then, we performed control experiments where the same black phosphorus flake is compressed along the AC and ZZ directions finding that, when compressed along the ZZ direction, the flake buckles forming ripples with a period about ~ 1.4 times longer that obtained for compression along the AC direction. Note that in previous buckling metrology measurements on other 2D materials, like graphene or transition metal dichalcogenides, the controlled compression aligned along well-defined crystal directions were not reported. A quantitative analysis of the black phosphorus buckling allows us to determine the Youngs's modulus of few-layer black phosphorus along the AC direction ($E_{bP\_AC} = 35.1 \pm 6.3$ GPa) and the ZZ direction ($E_{bP\_ZZ} = 93.3 \pm 21.8$ GPa).

Black phosphorus flakes have been buckled by transferring them onto a compliant elastomeric substrate that has been initially stretched. After the flake transfer the substrate pre-stress is released exerting a uniaxial compression on the black phosphorus flakes. As elastomeric material we employ Gel-Film (WF x4 6.0 mil by Gel-Pak) and we apply the pre-stress by bending it (See Figure 1a). Black phosphorus flakes are exfoliated from bulk black phosphorus crystals (HQ Graphene) with Nitto tape (Nitto Denko corp. SPV224) and directly transferred onto the surface of the stretched Gel-Film substrate. After the transfer of the flake, we release the stress by unbending the Gel-Film substrate, and the sample is inspected under an optical microscope to find the buckled black phosphorus flakes. Figure 1b shows a transmission mode optical microscopy image of a few-layer black phosphorus flake displaying a clear periodic ripple pattern that arises from the interplay between the flake buckling and the flake-substrate adhesion interaction.

Using a (linear) polarization analyser during the optical microscopy inspection one can easily determine the crystal orientation of the flakes due to the strong linear dichroism of black phosphorus. In fact, the optical transmittance of black phosphorus is larger for light linearly polarized along the ZZ crystal direction.[50–53] Figure 1c shows the dependence of the red, green and blue image channels transmittance (calculated as the light intensity transmitted though the black phosphorus flake divided by the light intensity transmitted though the bare substrate) as a function of the angle between the linear polarizer direction and the horizontal axis (*i.e.* 0° means polarization parallel to the horizontal axis and 90° means polarization parallel to the vertical axis). The angular dependence of the transmittance follows the Malus' law ($T \propto \cos^2(\theta+\delta\theta)$) as expected for a material with a strong linear dichroism. From the angle values where the transmittance reaches the maximum and the minimum one can determine the ZZ and AC directions of the black phosphorus flake respectively. These directions are displayed in Figure 1b making it possible to determine the relative orientation of the ripples with respect to the crystalline directions: the ripples are oriented almost parallel to the ZZ crystal orientation.

According to theoretical calculations, the Young's modulus of black phosphorus is expected to be ~3-4 times larger along the ZZ than along the AC [54–57] and thus it results energetically favourable to bend the black phosphorus lattice along the ZZ axis. Moreover, a lower Young's modulus along the AC direction also means that the compression stress needed to buckle the black phosphorus along that direction is also lower than that needed to buckle it along the ZZ direction.[42,43,58] These reasons explain why the black phosphorus flakes preferentially buckle forming ripples parallel to the ZZ direction. Figure 1c presents a polar histogram of the relative angle between the ZZ direction and the ripple orientation where a marked preferential alignment along the ZZ direction is seen.

Due to the preferential alignment of the ripples parallel to the ZZ direction, the Young's modulus of black phosphorus along the AC direction can be determined by a quantitative analysis of the buckling-induced rippling period for black phosphorus flakes with different thicknesses. Indeed, there is a linear relationship between the ripple period and the flake thickness given by:[42,43,58]

$$\lambda = 2\pi h \left[\frac{(1-v_s^2)E_f}{3(1-v_f^2)E_s}\right]^{1/3} \qquad [1]$$

where $h$ is the flake thickness, $v_s$ and $v_f$ are the Poisson's ratio of substrate and flake and $E_s$ and $E_f$ are the Young's modulus of the substrate and flake respectively. Therefore, provided that one knows the values of the Poisson's ratio of black phosphorus ($v_{bP\_AC}$ = 0.40, $v_{bP\_ZZ}$ = 0.93),[59] and substrate ($v_s$ = 0.5)[60] and the Young's modulus of the substrate $E_s$ = 492 ± 11 kPa,[45] the Young's modulus can be readily determined from the slope of the linear relationship between the ripple period and flake thickness:

$$E_{bP\_AC} = \frac{3(1-v_{bP\_AC} \cdot v_{bP\_ZZ})E_s}{8\pi^3(1-v_s^2)}\left(\frac{\lambda}{t}\right)^3 \qquad [2],$$

Figure 2a shows four examples of rippled flakes with different thicknesses (ranging from 9 nm to 23 nm) that display a marked thickness-dependent period of ripple. The thickness of the flakes has been determined through quantitative analysis of the transmission mode optical images of the flakes (see the Electronic Supporting Information Figures S1 and S2 for details about the thickness determination procedure). As the whole measurement is carried out by optical microscopy, one can determine both the thickness of the flakes and the period of the ripples very quickly right after the exfoliation of black phosphorus (samples are exposed to air less than a 1 minute from their exfoliation to their experimental characterization) ensuring that this method provides the mechanical properties of pristine (not environmentally degraded black phosphorus). Note that in previous works on the mechanical properties of black phosphorus the studied flakes have been exposed to air for longer periods (atomic force microscopy was used to locate the flakes and to make force-indentation measurements)[24,25,28] and in some cases the flakes have even been subjected to several wet chemistry steps or electron beam irradiation (involved in the fabrication of freely suspended black phosphorus beams).[24,28] We point the reader to the Electronic Supporting Information for a study about the role of environmental exposure on the buckling (and its in-plane anisotropy) of black phosphorus (Figures S3 and S4).

Figure 2b summarizes the results, acquired for 22 flakes with thicknesses ranging 9 nm to 23 nm, that follows a linear relationship with a slope of 192 ± 9 from which one can determine the Young's modulus along the AC direction $E_{bP\_AC}$ = 35.1 ± 6.3 GPa.

In order to get an insight about the anisotropic mechanical properties of black phosphorus we transfer black phosphorus flakes onto a flat (unstrained) Gel-Film substrate, we determine their thickness and crystal orientation by optical microscopy (as detailed above) and we subject the same flakes to compressive strain along the AC and ZZ directions by pinching the surface of the Gel-Film with two glass slides as illustrated in Figure 3a. We compressed the Gel-Film substrate until the black phosphorus flake just starts to buckle and we stop compressing it further at that point (the approximate compression value is ~10%). Note that due to the large Young's modulus mismatch between the Gel-Film and black phosphorus this compression would translate into a much lower compressive strain on the flake. Flakes with relatively large areas (~4000 μm$^2$) with homogeneous thickness are selected for this experiment to facilitate the analysis. Figure 3b shows an example where the same flake is subjected to compressive strain along the AC and ZZ directions yielding to almost perpendicular ripple pattern with a sizeable different period (see the comparison between two line-profiles, for strain along ZZ and AC directions, in Figure 3c). Figure 3d summarizes the measured period in 4 different flakes upon compression along their ZZ and AC directions. Using Equation [2], one can determine the Young's modulus of black phosphorus along the ZZ direction from the Young's modulus value along the AC direction and the ratio between the ripple period for compression along the AC and ZZ directions ($\lambda_{AC}/\lambda_{ZZ}$ =1.39 ± 0.15):

$$E_{bp\_ZZ} = E_{bp\_AC} \cdot \left(\frac{\lambda_{AC}}{\lambda_{ZZ}}\right)^3 \qquad [3]$$

We obtain $E_{bP\_ZZ}$ = 93.3 ± 21.8 GPa. Note that $E_{bP\_ZZ} \cdot \nu_{bP\_AC} \approx E_{bP\_AC} \cdot \nu_{bP\_ZZ}$, as expected from the symmetries of the compliance/stiffness tensor in the strain vs. stress relationship.

In order to compare our results with the ones reported in the literature, we have summarized in Table 1 both theoretical and experimental results from the literature. Our Young's modulus value along the AC direction is in a good agreement with the theoretical values reported in the literature that spans between 21-52 GPa.[54–56,61,62] However, the span of the theoretical values of the Young's modulus along the ZZ direction is much larger 91-192 GPa, being our experimental Young's modulus for the ZZ direction is compatible with the lower bound of these theoretical values.[54–56] Regarding the previously reported experimental works, we first have to note that in references [25,27] the measurements were carried out by atomic force microscopy (AFM) indentation on circular drumheads and therefore it was not possible to probe the direction dependence of the Young's modulus. In references [24,28], on the other hand, the measurements were carried out by AFM indentation on freely-suspended doubly clamped beams which allowed to extract the Young's modulus along the AC and ZZ directions. In these works, however, the reported values ($E_{bP\_AC}$ ~ 27 GPa and $E_{bP\_ZZ}$ ~ 58-65 GPa) are noticeably lower than our obtained values. We must note that in these previous works, during the fabrication of the freely-suspended beams the black phosphorus flakes were exposed to several steps of wet-chemistry and/or electron beam exposure which could effectively induce the degradation of black phosphorus thus affecting their mechanical properties. Alternatively to these nanoindentation works a recent work extracted the Young's modulus of a 95 nm thick black phosphorus of a circular drumhead mechanical resonator by comparing the frequencies and mode shapes of high order mechanical resonances.[26] Interestingly, this nanomechanical resonator measurement on a 95 nm thick black phosphorus flake provides Young's modulus values ($E_{bP\_AC}$ ~ 46 GPa and $E_{bP\_ZZ}$ ~ 116 GPa) close to our results (obtained by buckling-metrology method) for 9-23 nm thick flakes. Moreover, among the experimental results on few-layer black phosphorus the one in reference [26] and our results based on the buckling metrology method are the ones that provide Young's modulus closer to those obtained from ultrasound velocity measurements in bulk black phosphorus crystals ($E_{bP\_AC}$ ~ 55 GPa and $E_{bP\_ZZ}$ ~ 179 GPa). Figure 4 shows a graphical comparison between the different reported values of Young's modulus and our experimental values.

**Conclusions**

In summary, we subjected few-layer black phosphorus flakes, deposited onto compliant elastomeric substrates, to uniaxial compressive strain yielding to their buckling and inducing a periodic rippling. The quantitative analysis of the period of the ripples for flakes with different thickness allows us to determine the Young's modulus of black phosphorus along the AC and ZZ directions: $E_{bP\_AC}$ = 35.1 ± 6.3 GPa and $E_{bP\_ZZ}$ = 93.3 ± 21.8 GPa. These results provide an experimental evidence of the in-plane anisotropy of the intrinsic mechanical properties of pristine black phosphorus as the samples have been exposed to air less than 30-45 minutes, preventing their environmental degradation.

**Experimental**

**Materials**

Black phosphorus samples were prepared out of a bulk crystal (HQ Graphene®) by mechanical exfoliation with Nitto 224SPV tape (Nitto Denko®). The elastomer substrate used in this work is a commercially available polydimethylsiloxane-based substrate manufactured by Gel-Pak® (Gel-Film® WF X4 6.0 mil).

**Determination of the Young's modulus of the Gel-Film® substrate**

The Young's modulus of the elastomeric substrate has been previously determined in reference [29] by force vs. elongation experiments (see the Supporting Information of reference [29]).

**Optical microscopy**

Optical microscopy images have been acquired with an AM Scope BA MET310-T upright metallurgical microscope equipped with an AM Scope mu1803 camera with 18 megapixels. The calibration of the optical magnification system has been carried out by imaging standard samples: one CD, one DVD, one DVD-R, and two diffraction gratings with 300 lines/mm (Thorlabs GR13-0305) and 600 lines/mm (Thorlabs GR13-0605).

**Image analysis**

The quantitative analysis of the transmittance of the flakes and the rippling wavelength has been carried out using Gwyddion® software.[39]

**Thickness determination**

The thickness determination has been carried out by extracting the transmittance of the red, green and blue channel of the transmission mode optical microscopy images and comparing it with the results of reference (not-buckled) samples. See the Supporting Information for more details.

**Atomic Force Microscopy**

Atomic force microscopy measurements were carried out with an ezAFM from NanoMagnetics Instruments operated in tapping mode with cantilevers of 40 N/m and a resonance frequency of 300 kHz.

**Conflicts of interest**

There are no conflicts to declare.

**Acknowledgements**

This project has received funding from the European Research Council (ERC) under the European Union's Horizon 2020 research and innovation programme (grant agreement n° 755655, ERC-StG 2017 project 2D-TOPSENSE). R.F. acknowledges the support from the Spanish Ministry of Economy, Industry and Competitiveness through a Juan de la Cierva-formación fellowship 2017 FJCI-2017-32919

FIGURES:

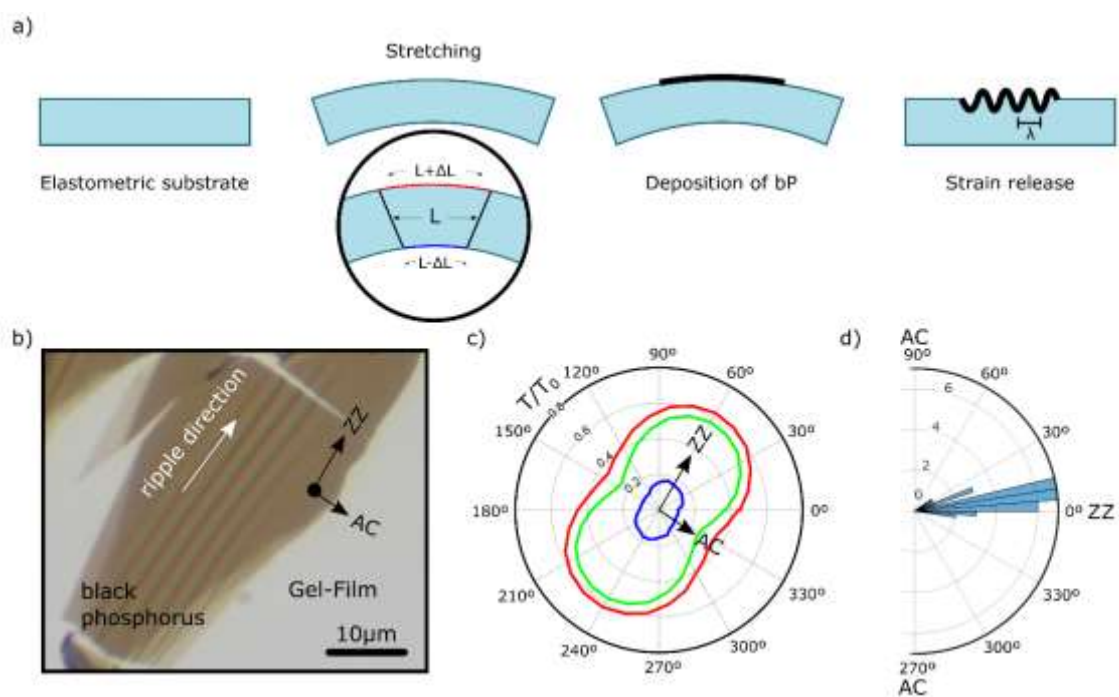

Figure 1. **a)** Schematic diagram of the process employed to fabricate the samples, where the flakes are transferred onto a stretched elastomeric substrate. When the stress is released, the flakes are subjected to compressive stress that produces ripples with a certain thickness-dependent period. **b)** Transmission mode optical microscopy images of a black phosphorus multilayer flake after releasing the stress on the elastomeric substrate, where the ripples are (almost) parallel to the zig-zag direction (ZZ) shown in c). **c)** Angular dependence of the optical transmission measured on a black phosphorus flake normalized to the transmission of the substrate, by varying the angle of the linearly polarized illumination and where each color correspond to the different channels (Red, Green and Blue). The maximum value corresponds to the ZZ direction and the minimum to the armchair direction (AC). **d)** Histogram of the difference between the angle of the ZZ and ripples directions for various samples.

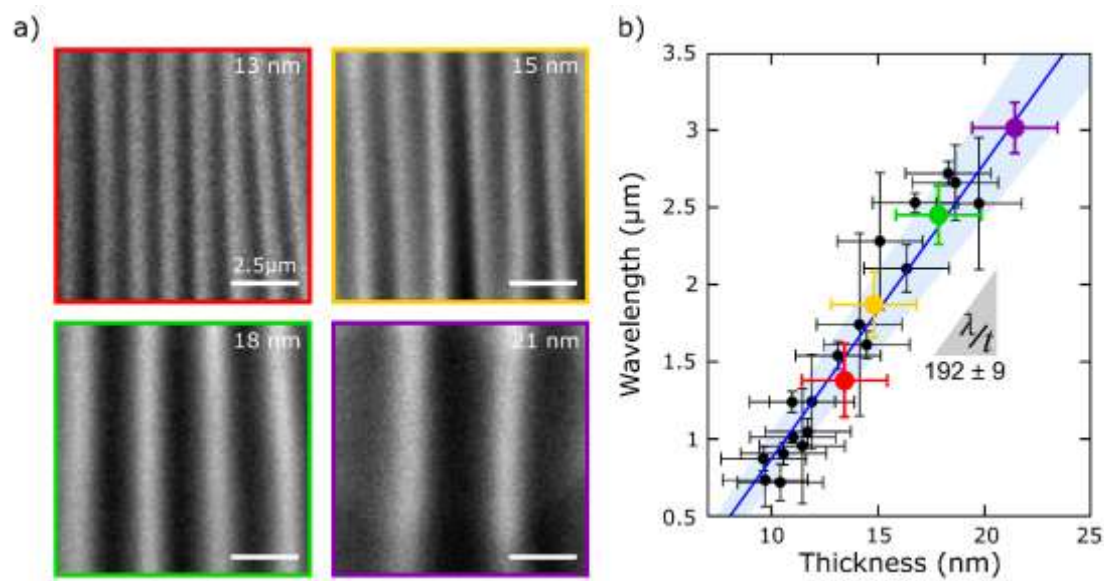

Figure 2. **a)** Optical microscopy images of ripples with different periods for 4 black phosphorus flakes of different thickness. **b)** Wavelength vs thickness graph for several black phosphorus flakes with different thickness, their error bars with 95% confidence curves. The solid blue line is the linear fit and the shaded area around indicates the uncertainty of the fit (95% confidence).

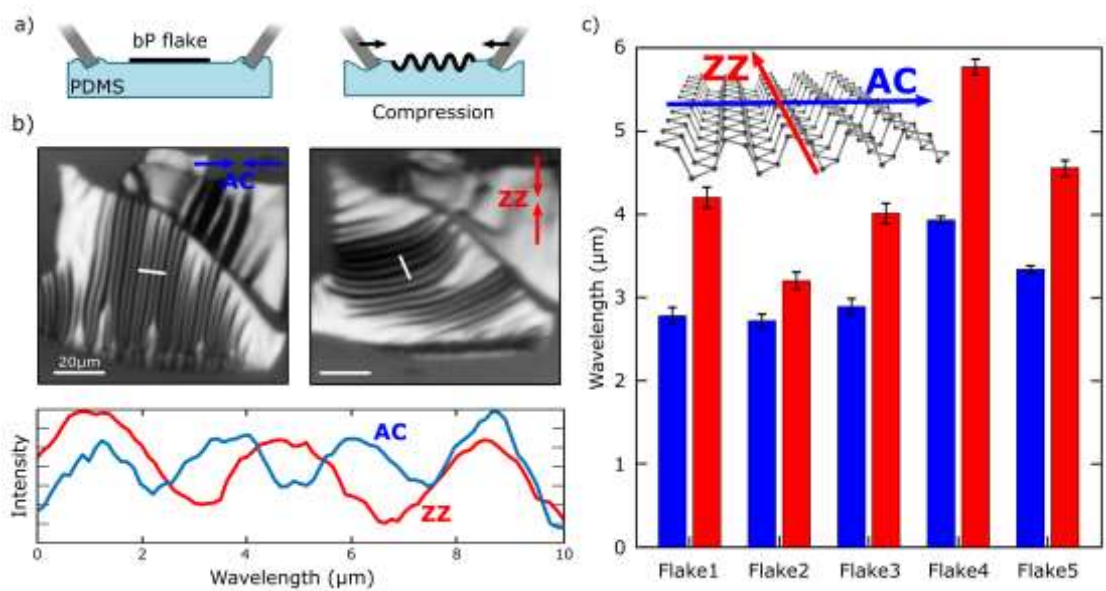

Figure 3. **a)** Process of ripples formation in the two directions (ZZ and AC) in a controlled way two manipulators to apply the compression along both directions. **b)** Optical microscopy images of black phosphorus compressed in both directions. The period of the ripple pattern sizeably depends on the direction where the compression is applied. The coloured arrows indicate the ZZ (blue) and AC (red) compression direction. **c)** Comparison between the ZZ and AC ripple period measured for different flakes.

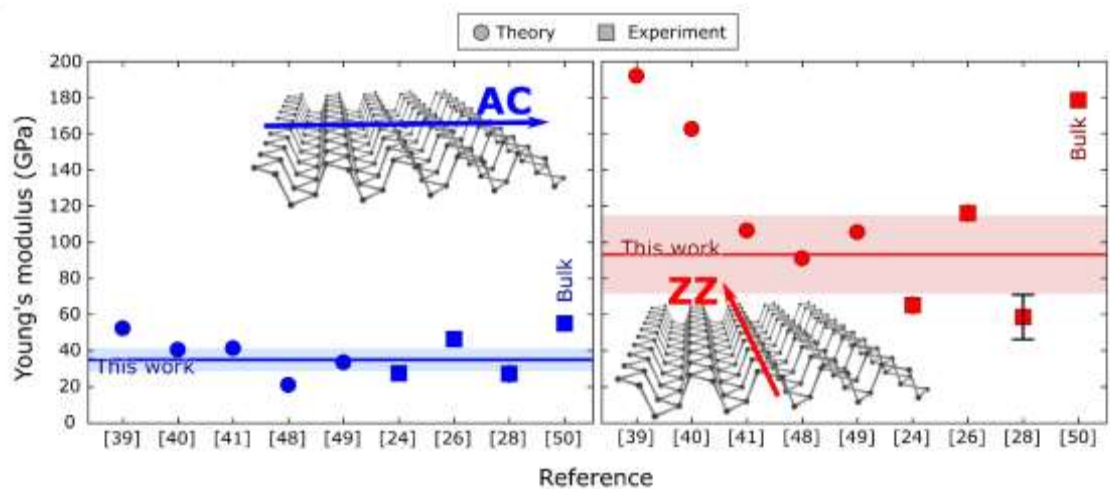

Figure 4. Graphical comparison between the reported theoretical and experimental values of the Young's modulus in the literature (symbols) and our experimental results (solid horizontal lines, the shaded area indicates the uncertainty). The values along the AC direction are displayed at the left and along the ZZ at the right.

Table 1. Summary of the reported valued (both theoretical and experimental) in the literature, indicating the method and conditions employed to obtain them.

| | Method and conditions (environment, thickness) | | *E* [GPa] | | Ref. |
|---|---|---|---|---|---|
| | | | AC | ZZ | |
| Theory | Density functional theory | | 41.3 | 106.4 | 56 |
| | Density functional theory | | 37-44 | 159-166 | 55 |
| | Density functional theory | | 52.3 | 191.9 | 54 |
| | Molecular dynamics | | 21 | 91 | 61 |
| | Molecular dynamics | | 33.5 | 105.5 | 62 |
| Experiment | AFM indentation | air, 15-25 nm | 27.2 ± 4.1 | 58.6 ± 11.7 | 28 |
| | AFM indentation | high vacuum, 4-30 nm | 46 ± 10 | | 27 |
| | Nanomechanical resonators | air, 95 nm | 46.5 ± 0.8 | 116.1 ± 1.9 | 26 |
| | AFM indentation | air, 14-34 nm | 89.7±26.4 to 276±32.4 | | 25 |
| | AFM indentation | air, 58-151 nm | 27.38±2.35 | 65.16±4.45 | 24 |
| | Ultrasound velocity | air, bulk | 55.1 | 178.6 | 63 |
| | Buckling-metrology | air, 9-23 nm | 35.1 ± 6.3 | 93.3 ± 21.8 | This work |

* This experimental measurement did not allow to resolve the Young's modulus for different crystal orientations.

# Electronic Supporting Information:

# Anisotropic buckling of few-layer black phosphorus

*Luis Vaquero-Garzon[1], Riccardo Frisenda[1], Andres Castellanos-Gomez[1]*

[1]Materials Science Factory. Instituto de Ciencia de Materiales de Madrid (ICMM-CSIC), Madrid, E-28049, Spain.

riccardo.frisenda@csic.es , andres.castellanos@csic.es

**Thickness determination through quantitative analysis of the transmission mode optical images of the flakes**

The determination of the thickness of the black phosphorus flakes studied in the main text has been made by quantitative analysis of the transmission mode optical images of these flakes.

In order to determine the thickness through optical microscopy we acquire transmission-mode microscopy images of several flakes with different thicknesses on the surface of a Gel-Film substrate. Then these flakes are transferred onto a $SiO_2$ substrate (performing AFM characterization on the Gel-Film substrate is rather challenging because of the large tip-Gel-Film adhesion force) and their thickness is measures with atomic force miscroscopy (AFM). All this process is summarized in Figure S1.

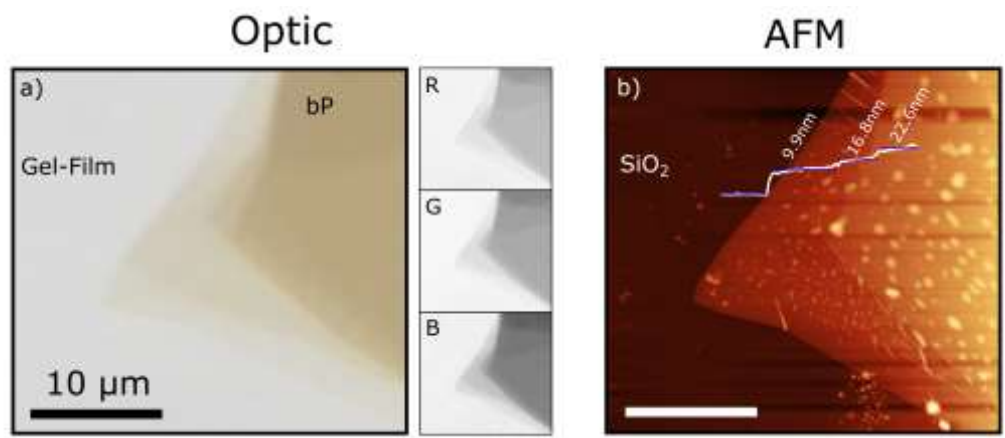

Figure S1. **a)** Optical microscopy image of multi-layer black phosphorus flake. Different RGB channels for the same flake. **b)** AFM image of the same flake in a) after being transferred onto a SiO$_2$/Si substrate. A topographic line profile over a region with some layers is included.

With the above data collected, a thickness versus normalized transmission graph can be made for the three channels (red R, green G, blue B) of the camera. Then, a polynomial fit to the data is used to obtain a relationship between thickness and transmission. The results are shown in the Figure S2.

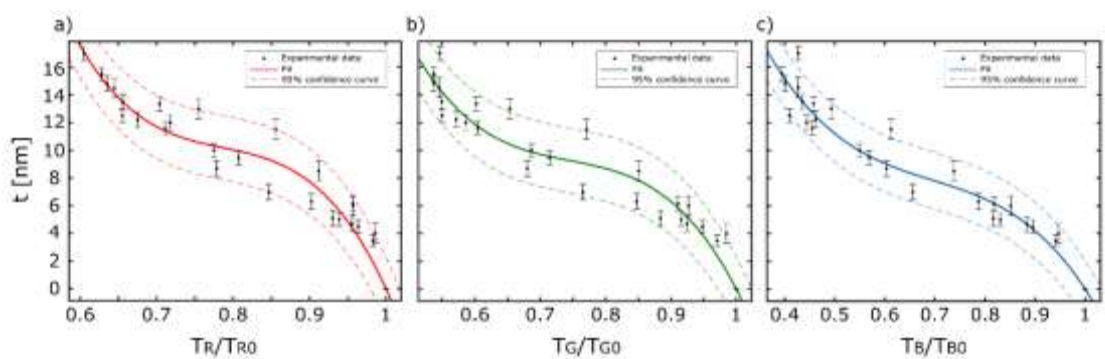

Figure S2. Thickness vs normalized optical transmission graph for some flakes and the polynomial fit with the 95% confidence curves for each channel **a)** Red **b)** Green **c)** Blue

By the graphs of Figure S2 we can obtain the thickness of a flake knowing the value of its normalized optical transmission, which makes this method quite comfortable. The error made by this approach is about ± 2 nm in the thickness.

## Effect of the environmental degradation on the buckling of black phosphorus

It has been thoroughly reported that black phosphorus tends to degrade upon environmental exposure by photo-induced oxidation.[1–3] This environmental degradation have been found to have an impact on the mechanical properties of black phosphorus but previous reports could not test the mechanical properties along different crystal orientations of the black phosphorus lattice.[4]

We have employed the buckling metrology method to test the effect of environmental exposure on the mechanical properties (and on its anisotropy) of black phosphorus. Figure S3(a) shows a sequence of transmission mode optical microscopy images of a buckled black phosphorus flake as a function of time. Figure S3(b) shows the line profile along the blue lines in (a) where the time evolution of the rippling can be easily observed. The degradation of the flake can be readily observed from the change in the flake transmittance as the generation of a phosphorus oxide crusts on the surface will reduce the optical absorption (while black phosphorus has a narrow gap, phosphorus oxides are wide band gap semiconductors). Figure S3(c) shows the line-profiles along the red lines in (a) where the time-dependent transmittance can be seen. Figure S3(d) summarizes the time evolution of the ripple wavelengths and transmittance measured in different spots of the sample.

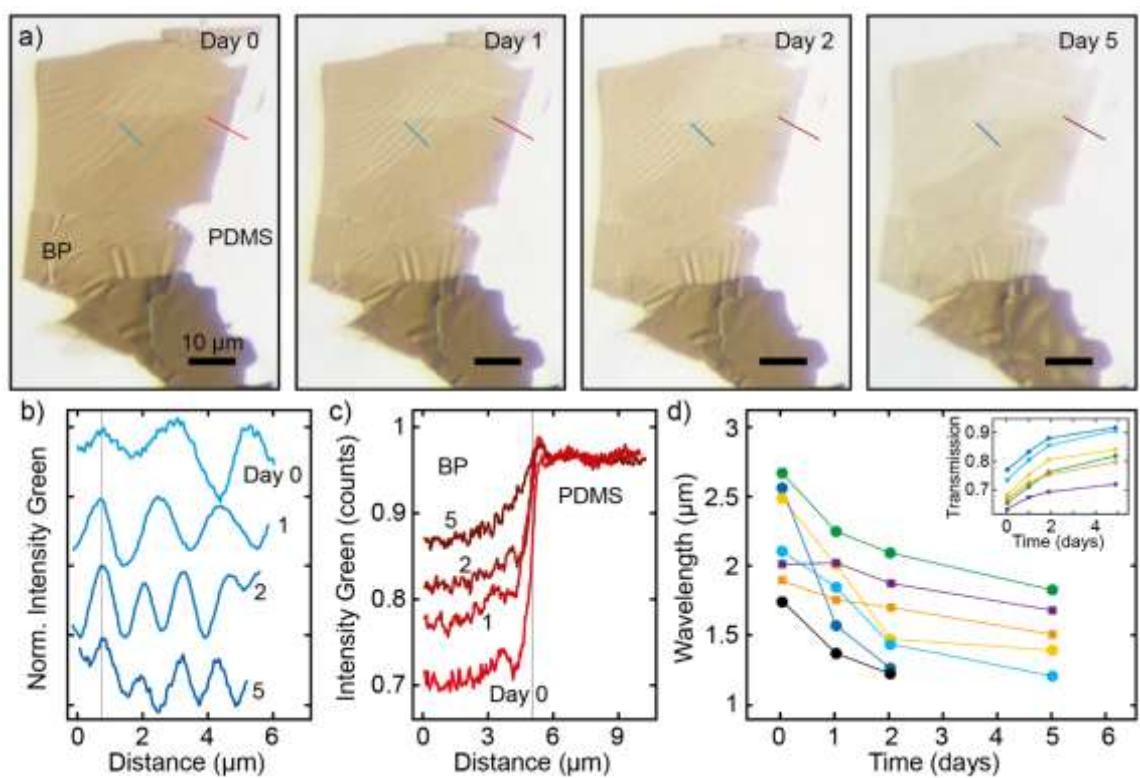

Figure S3. **a)** Sequence of transmission optical microscopy images of a buckled black phosphorus flake upon time exposure. **b)** Line profiles along buckling-induced ripples at different time exposures. **c)** Time evolution of the transmission of the black phosphorus flake. **d)** Summary of the time evolution of the ripple wavelength and the transmission (of the green channel) measured at different locations of the black phosphorus flake upon air exposure. Note: average relative humidity 20% during the experiment.

To get a deeper insight into the effect of the environmental induced degradation on the mechanical properties of black phosphorus we compare the results of the controlled buckling experiments (where the compression is applied in a controlled way along the ZZ and the AC directions) on pristine black phosphorus flakes and on aged flakes (after 9 days of air exposure). Figure S4(a) shows optical microscopy images of an aged black phosphorus flake (after 9 days of air exposure with an average relative humidity 20%) compressed along both AC (blue) and ZZ (red) directions. A direct comparison between the ripples profiles measured on a pristine flake and on the aged flake shown in Figure S4(a) is displayed in Figure S4(b). A summary of the anisotropy measured on pristine flakes and on two different aged black phosphorus flakes is shown in Figure S4(c).

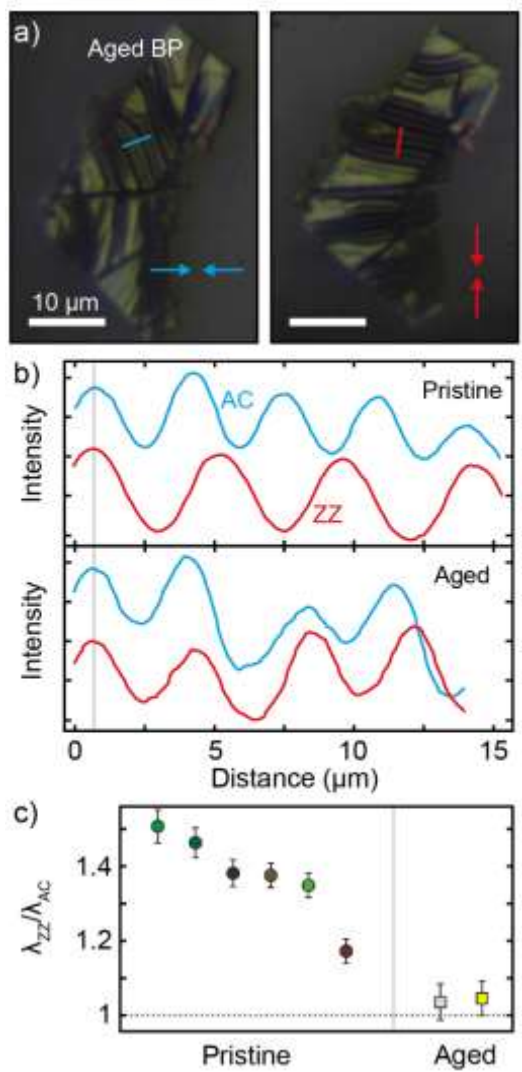

Figure S4. **a)** Optical microscopy images of an aged black phosphorus flake (after 9 days of air exposure with an average relative humidity 20%) compressed along both ZZ (red) and AC (blue) directions. **b)** Comparison between the ripples profiles measured on a pristine flake and on the aged flake. **c)** Comparison between the buckling anisotropy measured on pristine flakes and on two aged black phosphorus flakes.

## Electronic Supporting Information References: